# Limits of commutativity on abstract data types

Carmelo MALTA, José MARTINEZ

**Abstract**
We present some formal properties of (symmetrical) commutativity, the major criterion used in transactional systems, which allow us to fully understand its advantages and disadvantages. The main result is that commutativity is subject to the same limitation as compatibility for arbitrary objects. However, commutativity has also a number of attracting properties, one of which is related to recovery and, to our knowledge, has not been exploited in the literature. Advantages and disadvantages are illustrated on abstract data types of interest. We also show how limits of commutativity have been circumvented, which gives guidelines for doing so (or not!).

## 1. INTRODUCTION

In shared database systems, users access to data concurrently [Bernstein et al. 87]. The accesses are done inside a programming construct, named a *transaction*, which is a unit of consistency, i. e., a program which converts consistent data into consistent data [Gray 81]. To improve throughput, interleaved executions of transactions must be allowed; however, to enforce consistency, these interleaved executions must look like a serial execution. This syntactic criterion of consistency is called *serializability*.

Interactions between transactions take place through the use of common objects in the database. Originally, operations were merely uninterpreted reads and writes [Kedem & Silberschatz 83]; read is compatible with itself, while write is exclusive. However, *compatibility* was too strict a criterion for dealing with so-called "hot-spots" [O'Neil 86], thus an enhanced criterion was proposed: *commutativity*. This new one takes into account the semantics of operations on *abstract data types* (ADTs) in order to decrease the number of conflicts, which are in fact *pseudo-conflicts* [Schwarz & Spector 84]. Let us consider the DIRECTORY example to have an intuitive idea of the advantages of commutativity over compatibility.

A DIRECTORY is an ADT whose structure is a list of entries containing different names and associated items, say files in an operating system, and has a number of operations among which CREATE and DELETE. Both operations modify the directory, therefore both are writers and cannot be executed concurrently if compatibility is used. Commutativity allows a finer view of these operations, and, considering the name of the entry which is created or deleted, it can be concluded that CREATE and DELETE are in conflict only if

---

This work was supported in part by the PRCs BD3 and $C^3$ coordinated by the Centre National de la Recherche Scientifique (CNRS), and in part by the Ministère de la Recherche et de la Technologie (MRT).

they manipulate the same entry. Obviously, the number of conflicts is drastically reduced.

If advantages of commutativity have been clearly illustrated through examples, by contrast, possible limitations have not been investigated. Questions are: "Commutativity being certainly not a panacea, which are its limits? An then, where is it the most effective?" In order to answer them, we present some *folk-theorems* [Harel 80] related to commutativity of operations on ADTs. The main result is heavily indebted to our choice of formal approach: a functional model which highlights the important notions of domains and codomains.

The outline of this paper is as follows: First, we introduce our model and distinguish four conditions that functions on an ADT must satisfy. Next, we prove that for arbitrary objects, commutativity is not much more powerful than compatibility! However, we also prove attractive properties of commutativity. Two of them are well-known, but the last has practical implications of interest which have not been exploited in the literature. Finally, being aware of the limits of commutativity, we discuss the design of typical commutative ADTs. In particular, one can understand how some authors circumvent these limitations by using additional properties or weakening the required conditions of commutativity.

## 2. THE MODEL

In order to study properties of commutativity on ADTs, we use a functional formalism.

### 2.1. Operations as functions

Each operation on an ADT is expressed as one or many functions. For instance, let us consider the SET ADT and the INSERT(x) operation. We distinguish two functions: a first one defined from the set of SETs including x onto itself, and a second one defined from the sets which do not contain x to the sets including it. This differentiation is made clear in subsection 2.3. In practice, this point of view on operations has been implemented [Malta & Martinez 91b].

*definition 1*

Let $\mathbf{F} = (f_i)_{i=1}^{n}$ be a bag of functions defined as follows:

$f_i: A_i \to B_i$

We also define:

$A_f = \bigcap_{i=1}^{n} A_i$

$B_f = \bigcap_{i=1}^{n} B_i$

$\mathbf{F}$ is a bag of functions rather than a set because a function does not necessarily commute with itself. With a bag, self-commuting functions have just to be duplicated.

### 2.2. Composition (C1)

A bag of functions is intended to group functions which commute, therefore they must be composable in any order. To ensure this composition, a condition is required on the domains and codomains of the functions.

*definition 2*

Let $\mathbf{F}$ be a bag of pairwise commutative functions, then $\mathbf{F}$ must verify the composition condition (C1):

$\forall i: 1 \leq i \leq n,$
$\forall j: (1 \leq j \leq n) \wedge (j \neq i),$
$B_j \subseteq A_i$

### 2.3. Compensability (C2)

The effects of an operation may have to be undone, due to either a reject of the corresponding transaction, or a crash of the system. We then impose that each function have a left-inverse function.

*definition 3*
Let **F** be a bag of pairwise commutative functions, then we must have **F**$^{-1}$ = $(f_i^{-1})_{i=1}^{n}$ a bag of left-inverse functions:
    $f_i^{-1}$: $B_i \rightarrow A_i$
such that < **F**, **F**$^{-1}$ > verifies the compensability condition (C2):
    $\forall$ i: $1 \leq i \leq n$,
    $\forall$ x $\in A_i$,
      $f_i^{-1} \cdot f_i(x) = x$

This condition can be satisfied even for operations which simply write a new value, by restricting each $A_i$ to a single value.

### 2.4. State commutativity (C3)
The preceding conditions were either omitted, or implicit. The following definition introduces the first part of the explicited condition of commutativity.

*definition 4*
Let **F** be a bag of pairwise commutative functions, then **F** must satisfy the state commutativity condition (C3):
    $\forall$ i: $1 \leq i \leq n$,
    $\forall$ j: $(1 \leq j \leq n) \wedge (j \neq i)$,
    $\forall$ x $\in A_i \cap A_j$,
      $f_j \circ f_i(x) = f_i \circ f_j(x)$

We do not try to investigate the notion of state equivalence since equality is already an equivalence relation and consequently needs (just) to be defined adequately for each ADT (See [Weihl 88] for another definition).

### 2.5. View independence (C4)
In the previous definitions, only in-parameters of operations on ADTs are (implicitly) expressed. We explicitly represent out-parameters.

*definition 5*
Let **F** be a bag of pairwise commutative functions, then we also have **F**$_T$ = $(f_{T_i})_{i=1}^{n}$ another bag of functions defined as follows:
    $f_{T_i}$: $B_i \rightarrow E_i$

$f_{T_i} \circ f_i(x)$ is the view that the transaction T gets from the application of the operation associated to $f_i$ on the object x.

This model of out-parameters is consistent because each function has a left-inverse function, that is, $f_T$ can be defined, if there is no short cut, as $f'_T \circ f^{-1}$.

The last condition, second part of the familiar definition of commutativity, requires invariance of out-parameters. In the properties, C3 and C4 are never used together.

*definition 6*
Let **F** be a bag of pairwise commutative functions, then < **F**, **F**$_T$ > must verify the view independence condition (C4):
    $\forall$ i: $1 \leq i \leq n$,
    $\forall$ j: $(1 \leq j \leq n) \wedge (j \neq i)$,
    $\forall$ x $\in A_i \cap A_j$,
      $f_{T_i} \circ f_i \circ f_j(x) = f_{T_i} \circ f_i(x)$

Note that non-deterministic operations are not considered [Hesselink 88]. An extension to non-determinism is not straightforward: on the one hand, serializability can be lost; on the other hand, non-deterministic operations are unavoidable. If it can be shown that non-deterministic out-parameters do not change the behaviour of transactions, then all the results apply. This holds at the tuple-level of a multi-level transactional system: Creating a new tuple inserts it in some page and assigns it a tuple identifier which is not deterministic since it depends on the number of transactions creating or deleting tuples. Nonetheless, the behaviour of transactions is not affected by so low-level details.

**2.6. Remarks on the model**

Some remarks can be done about this formalization:

(i) Implicitly, each operation is executed on behalf of a different transaction. This simplifies the definitions. Moreover, a sequence of functions executed on behalf of a given transaction is reducible, by composition, to a unique function.

(ii) Also implicit is the fact that the commutativity that we study is conditional. In-parameters can generate as many functions as possible values, e. g., there is not just two INSERT functions as seen before, but an infinity of functions $\text{INSERT}_{x_1}$, $\text{INSERT}_{x_2}$, etc [Roesler & Burkhard 87]. Domains and codomains also generate different functions, i. e., in practice out-parameters are also exploited.

(iii) We define commutativity as being symmetrical. [Weihl 88] defines two kinds of commutativities: backward and forward. It turns out that the former is not symmetrical. The reason is that the bases for determining backward commutativity are not symmetrical. Nevertheless, with both commutativities, when two operations have been found commutative, they can be executed in any order.

(iv) Our formalism takes account of abstract data types solely. This is why we identify the notion of object with that of its value. Therefore, the results are directly applicable to multi-level transactions on a level-by-level basis, where objects are composed of objects of the underlying level [Cart et al. 90]. In contrast, they cannot be applied, as they are, to objects composed of references to other objects, e. g., instances of classes in object-oriented environments.

(v) The bags that we defined are conflict-preserving-serializable and characteristics of strict executions (i. e., conflicting operations are delayed). However, [Yannakakis 84] shows that the class of conflict-preserving-serializability is exactly the class of view-serializability plus a property of monotonicity, i. e., translated in our model, any sub-bag of a given bag verifying the four conditions, still satisfies them.

**3. PROPERTIES OF COMMUTATIVITY**

The following two subsections present respectively the advantages and disadvantages of commutativity as a criterion for concurrent accesses to shared data. First, disadvantages are highlighted; they concern parallelism which is much less increased, in general, than the profusion of the literature on this topic can lead one to suppose. Secondly, advantages are introduced; they chiefly concern recovery which can be simpler than implemented in some systems.

**3.1. Limits for concurrency**

We begin by some introductory and technical lemmas derived from the constraint on the domains and codomains of commutative functions, i. e., condition C1.

*lemma 1*
Let $\mathbf{F}$ verify C1, with $n \geq 2$, then:
$\quad \forall\ i:\ 1 \leq i \leq n,$
$\quad\quad B_i \subseteq A_{\mathbf{F}\text{-}(f_i)}$

*proof*
Obvious by definition of C1.       □

*lemma 2*
Let f verify C1, with $n \geq 2$, then:
$\quad \forall\ i:\ 1 \leq i \leq n,$
$\quad \forall\ j:\ (1 \leq j \leq n) \land (j \neq i),$
$\quad\quad B_i \cap B_j \subseteq A_{\mathbf{F}}$

*proof*
$B_i \subseteq A_{F-(f_i)}$ and $B_j \subseteq A_{F-(f_j)}$ [by lemma 1] then $B_i \cap B_j \subseteq A_{F-(f_i)} \cap A_{F-(f_j)}$. □

These lemmas lead directly to a first and simple theorem.

*theorem 1*
Let **F** verify C1, with $n \geq 2$, then:
$B_F \subseteq A_F$

*proof*
Obvious from lemma 2. □

Theorem 1 summarizes that (the necessary) condition C1 generates a close and very restrictive relationship between the sets of domains and codomains of a bag of pairwise commutative functions. This strong structure is not at all surprising as soon as we realize that it is based on set intersections and inclusions. Although this theorem is very simple, it is the unavoidable reason which makes commutativity so weak an enhancement for *arbitrary* objects. By itself, it has no more intrinsic interest, but we will see its implications connected to the following theorem.

In order to demonstrate theorem 2, we need a fundamental lemma of commutativity which takes into account state commutativity (condition C3). The full extension is part of the next subsection because it is a well-known advantage of commutativity.

*preliminary definition*
$\pi(\{1,...,n\})$ is the set of permutations of $\{1,...,n\}$.

*lemma 3*
Let **F** verify C1 and C3, then:
$\forall i: 1 \leq i \leq n$,
$\forall (k_1,...,k_{n-1}) \in \pi(\{1,...,n\} - \{i\})$,
$\forall x \in A_F$,
$\alpha \circ f_i(x) = f_i \circ \alpha(x)$
with $\alpha = f_{k_{n-1}} \circ ... \circ f_{k_1}$

*proof*
By induction on n.
Basis: Trivial for $n = 1$.
Induction step: $\alpha \circ f_i(x) = \beta \circ f_j \circ f_i(x) = \beta \circ f_i \circ f_j(x)$ [by C3], $A_F \subseteq A_j$, $f_j(x) \in B_j$, and $B_j \subseteq A_{F-(f_j)}$ [by lemma 1], then $\beta \circ f_i(f_j(x)) = f_i \circ \beta(f_j(x))$ [by induction] $= f_i \circ \alpha(x)$. □

Then, theorem 2 simply states that the final value resulting from the application of a bag of pairwise commutative functions is in $B_F$.

*theorem 2*
Let **F** verify C1 and C3, then:
$\forall x \in A_F$,
$f_n \circ ... \circ f_1(x) \in B_F$

*proof*
By induction on n.
Basis: By definition for $n = 1$.
Let **F** $= (f_1, f_2)$, then $f_1(f_2(x)) \in B_2$, $f_2(f_1(x)) \in B_1$, and $f_1 \circ f_2(x) = f_2 \circ f_1(x)$ [by C3], hence $f_1 \circ f_2(x) \in B_1 \cap B_2$.
Induction step: $f_n \circ ... \circ f_1(x) = f_n \circ \alpha(x) = \alpha \circ f_n(x)$ [by lemma 3], then, using the same reasoning as for the basis and applying the induction hypothesis, $\alpha(x) \in B_{F-(f_n)}$, and $B_{F-(f_n)} \subseteq A_F \subseteq A_n$ [by lemma 2], therefore $f_n(\alpha(x)) \in B_n$.
Conversely, $f_n(x) \in B_n$, and $B_n \subseteq A_{F-(f_n)}$ [by lemma 1], then $\alpha(f_n(x)) \in B_{F-(f_n)}$.
Consequently, $f_n \circ \alpha(x) \in B_{F-(f_n)} \cap B_n$. □

We are now ready to discuss all the (bad) implications of theorem 2. It needs two interpretations: a first one when $n \geq 2$, and a second one when $n = 1$. They have a common characteristic: *both* restrict concurrency.

When at least two functions are executed concurrently on the same object (i. e., when $n \geq 2$), theorem 1 holds too. Therefore, we deduce that commutativity of a bag of functions implies *convergence*, or at most monotonicity, in the sequence of values of an object since the initial value must be in $A_f$ and the final value is consequently in $B_f$, a subset. What is counter-intuitive in this proposition is that the views play no role; it would have been easier to understand that whenever an operation returns a view of an object, this snapshot limits further modifications.

Only when there is a unique transaction (i. e., when $n = 1$) can the value of an object be completely modified because condition C1 is not effective. In other words, a function which domain and codomain do not intersect, is always exclusive of any other! This is a corollary of theorem 1.

*corollary 1*
Let **F** verify C1 and C3, and let $f \in$ **F** be:
  $f: A \rightarrow B$
such that $A \cap B = \emptyset$, then:
  $A_\mathbf{F} \neq \emptyset \Rightarrow |\mathbf{F}| = 1$

*proof*
Let us proceed by contradiction and choose $f \in$ **F** such that $|\mathbf{F}| > 1$ and $A \cap B = \emptyset$, then by definition $A_\mathbf{F} \subseteq A$ and $B_\mathbf{F} \subseteq B$, which clearly contradicts theorem 1, except if $A_\mathbf{F} = \emptyset$.     □

Condition C1 and corollary 1 serve to detect all the pairs of functions which *cannot* commute, independently of their semantics. They are especially helpful because the number of functions is greater than the number of operations and that each function is generally exclusive of its counterparts.

As an example, let us develop the STACK ADT in its entirety. We distinguish nine functions associated to four operations: EMPTY, CLEAR, POP, and PUSH. These functions are defined over the sets $\mathbf{S}^0$ and $\mathbf{S}^+$ representing respectively the set consisting of the empty stack and the set of non-empty stacks:

  EMPTY$_{yes}$:     $\mathbf{S}^0 \varnothing \mathbf{S}^0$
  EMPTY$_{no}$:      $\mathbf{S}^+ \varnothing \mathbf{S}^+$
  CLEAR$_{yes}$:     $\mathbf{S}^+ \varnothing \mathbf{S}^0$
  CLEAR$_{already}$: $\mathbf{S}^0 \varnothing \mathbf{S}^0$
  POP$_{empty}$:     $\mathbf{S}^0 \varnothing \mathbf{S}^0$
  POP$_{last}$:      $\mathbf{S}^+ \varnothing \mathbf{S}^0$
  POP$_{yes}$:       $\mathbf{S}^+ \varnothing \mathbf{S}^+$
  PUSH$_{first}$:    $\mathbf{S}^0 \varnothing \mathbf{S}^+$
  PUSH$_{yes}$:      $\mathbf{S}^+ \varnothing \mathbf{S}^+$

The reasons for mapping four operations into nine functions are two-fold: First, condition C2 imposes unique inverse functions, e. g., CLEAR$_{already}$ and CLEAR$_{yes}$ have different inverse functions; Secondly, some differentiations improve commutativity, e. g., if PUSH$_{first}$ and PUSH$_{yes}$ were not distinguished, the unique PUSH function would commute with no other function.

In Table 1, all the pairs of functions which do not verify condition C1 are immediately marked with *i* (for "impossible"). Next, there are three exclusive functions, CLEAR$_{yes}$, POP$_{last}$, and PUSH$_{first}$, which are marked with *i'*. At last, the couple POP$_{last}$/PUSH$_{first}$ is marked *i''* since the intersection of their domains, $\mathbf{S}^+$ and $\mathbf{S}^0$, is empty. Thus, the set of pairs of functions which have to be taken into consideration for potential commutativity is dramatically reduced, which simplifies the effective work of the designer.

This kind of tool to handle complex commutativity conditions can be used in conjunction with other methodologies [Roesler & Burkhard 87] [Chrysanthis et al. 91].

We saw in this subsection what is the major disadvantage of commutativity: a convergence phenomenon. The restriction to bounded sets, a synonym for "computer sets", gives a still worse result: $A_F$ becomes equal to $B_F$ as soon as there is three functions in $F$ [Martinez 92].

However, the introductory DIRECTORY example shows that convergence can be slow for some operations on some ADTs. Therefore, it is worth considering the advantages of commutativity.

### 3.2. Advantages for recovery

Among the advantages of commutativity, theorems 3 and 4 are real folk-theorems.

Theorem 3 states that a bag of pairwise commutative functions can be composed in any order without changing the final state of the object. In simpler words, a pairwise commutative relation is also transitive, (regardless of reflexivity).

*theorem 3*
Let $F$ verify C1 and C3, then:
$\forall (k_1,...,k_n) \in \pi(\{1,...,n\})$,
$\forall x \in A_F$,
$f_{k_n} \circ ... \circ f_{k_1}(x) = f_n \circ ... \circ f_1(x)$

*proof*
By induction on n.
Basis: Trivial for n = 1 and by definition for n = 2.
Induction step: $f_{k_n} \circ ... \circ f_{k_i} \circ ... \circ f_{k_1}(x) = \alpha \circ f_n \circ \beta(x) = f_n \circ \alpha \circ \beta(x)$ [by lemma 3] $= f_n \circ f_{n-1} \circ ... \circ f_1(x)$ [by induction]. □

Theorem 4 expresses that out-parameters are not sensitive to the order of application of commutative functions, and more accurately that out-parameters are not sensitive to whether commutative functions are applied or not. Then, theorem 4 can be seen as the second part of theorem 3, just as condition C4 can be considered the second part of condition C3. But theorem 4 also supports the use of commutativity with optimistic methods where concurrent operations of other transactions are not reflected on the workspace of a given transaction.

|     | $E_y$ | $E_n$ | $C_y$ | $C_a$ | $P_e$ | $P_l$ | $P_y$ | $U_f$ | $U_y$ |
|-----|-------|-------|-------|-------|-------|-------|-------|-------|-------|
| $E_y$ |   | i | i' |   |   | i' | i | i' | i |
| $E_n$ | i |   | i' | i | I | i' |   | i' |   |
| $C_y$ | i' | i' | i' | i' | i' | i' | i' | i' | i' |
| $C_a$ |   | i | i' |   |   | i' | i | i' | i |
| $P_e$ |   | i | i' |   |   | i' | i | i' | i |
| $P_l$ | i' | i' | i' | i' | i' | i' | i' | i'' | i' |
| $P_y$ | i |   | i' | i | I | i' |   | i' |   |
| $U_f$ | i' | i' | i' | i' | i' | i'' | i' | i' | i' |
| $U_y$ | i |   | i' | i | I | i' |   | i' |   |

**Table 1**: non-commutativity matrix for the STACK ADT

Another time, we can say in a simpler way that commutativity guarantees isolation of transactions.

*theorem 4*
Let $<F, F_T>$ verify C1, C3, and C4, then:
$\forall i: 1 \leq i \leq n$,
$\forall x \in A_F$,
$f_{T_i} \circ f_n \cdot ... \circ f_1(x) = f_{T_i} \circ f_i(x)$

*proof*
By induction on n.
Basis: Trivial for n = 1 and by definition for n = 2.
Induction step: $f_n \circ ... \circ f_1(x) = f_i \circ f_j \circ \alpha(x)$ [by theorem 1], $\alpha(x) \in B_{F-(f_i,f_j)}$ [by theorem 2], and $B_{F-(f_i,f_j)} \subseteq A_i \cap A_j$ [by C1 or lemma 2], therefore $f_{T_i} \circ f_i \circ f_j(\alpha(x)) =$

$f_{T_i} \circ f_i(\alpha(x))$ [by C4] = $f_{T_i} \circ f_i(x)$ [by induction].   □

Finally, here comes the major advantage of commutativity: Composing a bag of pairwise commutative functions with a subbag of its bag of inverse functions, where each inverse function is applied after its direct one, is equivalent to a composition where the undone functions were never executed.

*theorem 5*
Let $<\mathbf{F}, \mathbf{F}^{-1}>$ verify C1, C2, and C3, and let f' = $(f_1, ..., f_n, f_{n+1}, ..., f_{n+m})$ be such that f = $(f_1, ..., f_n)$, and, without loss of generality, $(f_{n+1}, ..., f_{n+m}) = (f_1^{-1}, ..., f_m^{-1})$ with $m \leq n$, then:
$\forall (k_1, ..., k_{n+m}) \in \pi(\{1, ..., n+m\})$,
$\forall x \in A_{\mathbf{F}}$,
( $\forall i: 1 < i \leq n + m$,
$k_i \geq n + 1 \Rightarrow$
$\exists j: 1 \leq j < i \mid k_i = n + k_j$ ) $\Rightarrow$
$f_{k_{n+m}} \circ ... \circ f_{k_1}(x) = f_n \circ ... \circ f_{m+1}(x)$
with improper notation when m = n.

*proof*
By induction on m.
Basis: By theorem 3 for m = 0.
Induction step: Let us take the minimal i such that $k_i \geq n + 1$, then there exists j < i such that $k_i = n + k_j$; let $k_j$ be l, then
$f_{k_{n+m}} \circ ... \circ f_{k_i} \circ ... \circ f_{k_j} \circ ... \circ f_{k_1}(x)$ =
$\alpha \circ f_l^{-1} \circ \beta \circ f_l \circ \gamma(x)$ =
$\alpha \circ f_l^{-1} \circ f_l \circ \beta \circ \gamma(x)$ [by theorem 3] =
$\alpha \circ \beta \circ \gamma(x) = f_n \circ ... \circ f_{m+1}(x)$ [by induction].   □

More simply, any inverse function can be applied at any moment after the application of its associated direct one. Moreover, theorem 5 states that no control is necessary between direct and inverse functions, nor between inverse functions.

Surprisingly, this theorem seems not to be exploited in literature. [Weikum 91] argues that unresolvable deadlocks can occur during a reject process if an inverse operation is less commutative than its direct operation. The solution implemented by [Brössler & Freisleben 89] is to make the direct and inverse operations have the same restrictions, i. e., two operations commute if and only if they commute and commute with the inverse function of each other. [Moss et al. 86] concludes that the issue of knowing if whenever two operations commute, their inverse operations also commute, should be addressed. [Weikum 91] conjectures that "it is always possible to design inverse actions with a conflict relation that is no more restrictive than that of their primary actions."

Theorem 5 establishes the fact that inverse operations need just to be atomic, which can be obtained with short-term locking for instance, and do not necessitate their own synchronization mechanism since they rely on commutativity of their direct operations. Then, inverse operation should not be treated uniformely as direct operations.

Note that this theorem does not imply, as a corollary, that inverse functions commute if direct ones do so. The reason is simply that condition C1 is generally not satisfied for inverse functions. However, it is satisfied in the subcase of bounded sets [Martinez 92].

### 3.3. Discussion
We have proved that commutativity has practical advantages of interest, (especially for rejecting operations), which increase parallelism both between in-progress and rejected transactions. However, we have also shown that commutativity is (just) super-compatibility and suffers the same drawbacks: a write access was exclusive with compatibility, exclusive operations are not eliminated by commutativity; read accesses could not modify the value of the shared object, commutative operations cannot modify the predicate describing the possible values of the object.

Several means have been exploited to allow finer concurrency: independence, non-determinism, mathematical commutativity of numbers, and relative recoverability.

*independence*

ADTs which dispose of great independence are SET, BAG, or MAP. Of this kind of objects are also the MAIL ADT in an operating system which utilizes a system-wide REGISTRY ADT of current users, or the introductory DIRECTORY ADT. As can be guessed, all these examples are instances of the general RELATION ADT.

Returning to the introductory example, it is not obvious to convince someone that commutativity is restrictive. In fact, the DIRECTORY ADT does not really take advantage of commutativity: it is just compatible! To prove this, consider the following implementation of the SET ADT: An instance is merely the characteristic function, i. e., an array of booleans, and the locking granule is the size of a boolean, then DELETE and INSERT are effectively compatible when applied to distinct items. Operations EMPTY or SIZE can be considered as macros.

*non-determinism*

Nevertheless, it is worth trying to circumvent these limits for less independent objects. For instance, [Schwarz & Spector 84] introduces a new ADT, the SEMIQUEUE, derived from a very constrained one, the FIFOQUEUE. A SEMIQUEUE has a weakened GET operation: it does not necessarily remove the oldest item in the queue but one of the oldest, i. e., fairness is imposed but not strict ordering. Therefore GET becomes a non-deterministic operation. Note that some independence has been introduced between the items of a SEMIQUEUE: the order relationship has been removed.

*mathematical commutativity*

There exists a very commutative ADT: the COUNTER. The original method is known as the *Escrow method* [O'Neil 86] and uses the mathematical property of commutativity on integer and real numbers. Maximizing concurrency on this kind of object requires to use either its state, or the set of active operations. But this introduces problems to decide whether an operation should be restarted or not [Ng 89].

*relative recoverability*

The STACK, (as well as the FIFOQUEUE), is an example of an ADT which cannot take advantage of commutativity on a great extent, even when associating several functions to each operation, as done with PUSH. For that kind of objects, the criterion of *relative recoverability* introduced by [Badrinath & Ramamritham 87] allows much more parallelism [Badrinath & Ramamritham 90], at the expense of theorems 3 and 4, however. This is because conditions C1 and C4 are weakened and C3 is no longer required.

As can be seen, high concurrency is always achieved either by taking advantage of natural additional properties, or by weakening the conditions imposed by pure commutativity. But what happen to arbitrary objects?

The most common type constructor is the tuple constructor. There exist strong dependencies between the fields of a tuple.

Consider an ADDRESS, composed of a number, a street, a ZIP, and a city. Someone can move house in the same city, or even the same street, but that is rather the exception. The ADDRESS ADT is composed of tightly coupled components. Tuple-based objects implying a strong dependency between the different fields, commutativity of operations can be simply deduced from commutativity, or even compatibility, of accesses to each fields of the tuple. We left this issue open in the domain of object-oriented systems [Malta & Martinez 91a]. But that is typically the case for classes whose structure is almost every time tuples (only construct in ORION [Banerjee et al. 87] and GemStone [Maier et al. 86], **tuple of** in O2 [Lécluse et al. 88]). Classes and methods being frequently added, removed or modified, the inherent limits of commutativity convince us that a very simple analysis of commutativity between methods should give as good results, if not better in terms of incurred overhead, as some very complicated technique.

Consequently, for tuple-based types, we recommend to rely on techniques even simpler than the one proposed by [Badrinath & Ramamritham 88].

## 4. CONCLUSION

Commutativity, the main criterion to control concurrent accesses to shared data in transactional systems, has been the subject of a big deal of papers. Illustrated with popular examples, it seems to be a great enhancement over compatibility.

The main result of this paper is to show that commutativity is subject to a convergence phenomenon which resembles the behaviour of compatibility: a write access is exclusive with compatibility, and exclusive operations are not eliminated by commutativity; read accesses cannot modify the value of the shared object, and commutative operations cannot weaken the predicate describing the set to which the object pertain.

Having in mind this limitation, we rapidly survey the techniques which have been used to enhance concurrency. This gives us guidelines for the design of concurrent abstract data types. The rules that we recommend to follow are:

- to use the full power of commutativity for independent objects only, i. e., objects for which convergence can be limited to subparts;
- to rely on very simple techniques, even based on compatibility, for tuple-based objects;
- to include a COUNTER ADT for dealing with "hot-spots."

On the good side, we prove that commutativity has nice properties, in particular for recovery, which both simplifies commutativity conditions, and eliminate the overhead of having to control concurrent accesses of inverse operations.

Also, the formal model gives an idea to help in determining *non*-commutativity of operations.

**Acknowledgments**
We wish to thank Michèle Cart, Jean Ferrié, and Jean-François Pons for helping us to improve the outline of this paper. Also, we sincerely acknowledge the careful readings and comments of the referees.

## 5. REFERENCES

[Badrinath & Ramamritham 87] Badrinath, B. R., Ramamritham, K.; *Semantics-based concurrency control: beyond commutativity*; Proceedings of the 3rd IEEE Int. Conf. on Data Engineering, Los Angeles, USA, February 1987

[Badrinath & Ramamritham 88] Badrinath, B. R., Ramamritham, K.; *Synchronizing transactions on objects*; IEEE Transactions On Computers, vol. 37, n° 5, May 1988, pp. 541-547 [Badrinath & Ramamritham 90] Badrinath, B. R., Ramamritham, K.; *Performance evaluation of semantics-based multilevel concurrency control protocols*; Proceedings of the 1990 Int. Conf. on the Management Of Data, Atlantic City, NJ, USA, May 1990, pp. 163-172

[Banerjee et al. 87] Banerjee, J., Chou, H.-T., Garza, J. F., Kim, W., Ballou, D. W. N., Kim, H.-J.; *Data model issues for object-oriented applications*; ACM Transactions On Information Systems, vol. 5, n° 1, January 1987, pp. 3-26

[Bernstein et al. 87] Bernstein, P. A., Hadzilacos, V., Goodman, N.; *Concurrency control and recovery in database systems*; Addison-Wesley Publishing Company, Reading, Massachusets, 1987

[Brössler & Freisleben 89] Brössler, P., Freisleben, B.; *Transactions on persistent objects*; Proceedings of the Workshop on Persistent Object Systems: their Design, Implementation, and Use, Newcastle, Australia, January 1989, pp. 19-35

[Cart et al. 90] Cart, M., Ferrié, J., Pons, J.-F.; *Objects modeling when using a multi-level transaction model*; Proceedings of the ECOOP/OOPSLA Workshop on Transactions and Objects, Ottawa, Canada, October 1990

[Chrysanthis et al. 91] Chrysanthis, P. K., Raghuram, S., Ramamritham, K.; *Extracting concurrency from objects: A methodology*; Proceedings of the Int. Conf. on the Management Of Data, Denver, Colorado, USA, May 1991, pp. 108-117

[Gray 81] Gray, J. N.; *The transaction concept: virtues and limitations*; Proceedings of the 7th Int. Conf. on Very Large Data Bases, Cannes, France, 1981, pp. 144-154

[Harel 80] Harel, D.; *On folk theorems*; Communications of the ACM, vol. 23, n° 7, July 1980, pp. 379-389

[Hesselink 88] Hesselink, W. H.; *A mathematical approach to nondeterminism in data types*; ACM Transactions On Programming Languages and Systems, vol. 10, n° 1, January 1988, pp. 87-117

[Kedem & Silberschatz 83] Kedem, Z. M., Silberschatz, A.; *Locking protocols: from exclusive to shared locks*; Journal of the ACM, vol. 30, n° 4, October 1983, pp. 787-804

[Lécluse et al. 88] Lécluse, C., Richard, P., Velez, F.; *O2, an object-oriented data model*; Proceedings of the ACM Int. Conf. on the Management Of Data, Chicago, Illinois, USA, June 1988

[Maier et al. 86] Maier, D., Stein, J., Otis, A., Purdy, A.; *Development of an object-oriented DBMS*; Proceedings of the Int. Conf. on Object-Oriented Programming Systems, Languages and Applications, September 1986

[Malta & Martinez 91a] Malta, C., Martinez, J.; *Controlling concurrent accesses in an object-oriented environment*; Proceedings of the 2nd Int. Symposium on Database Systems For Advanced Applications, Tokyo, Japan, April 1991, pp. 192-200

[Malta & Martinez 91b] Malta, C., Martinez, J.; *A framework for designing concurrent and recoverable abstract data types based on commutativity*; Proceedings of the 6th Int. Symposium on Computer and Information Sciences, Antalya, Side, Turkey, October 1991

[Martinez 92] Martinez, J.; *Contribution à la formalisation des problèmes de contrôle de concurrence et de recouvrement dans les bases de données à objets*; Ph. D. Thesis, September 1992, Université des Sciences et Techniques du

Languedoc, Montpellier, France, 160 p. (in French)

[Moss et al. 86] Moss, J. E. B., Griffith, N. D., Graham, M. H.; *Abstraction in recovery management*; Proceedings of the ACM Int. Conf. on the Management Of Data, Washington D.C., USA, May 1986, pp. 72-83

[Ng 89] Ng, T. P.; *Using histories to implement atomic objects*; ACM Transactions On Computer Systems, vol. 7, n° 4, November 1989, pp. 360-393

[O'Neil 86] O'Neil, P. E.; *The Escrow transactional method*; ACM Transactions On Database Systems, vol. 11, n° 4, December 1986, pp. 405-430

[Roesler & Burkhard 87] Roesler, M., Burkhard, W. A.; *Concurrency control scheme for shared objects: A peephole approach based on semantics*; Proceedings of the 7th Int. Conf. on Distributed Computing Systems, Berlin, West Germany, September 1987, pp. 224-231

[Schwarz & Spector 84] Schwarz, P. M., Spector, A. Z.; *Synchronizing shared abstract types*; ACM Transactions On Computer Systems, vol. 2, n° 3, August 1984, pp. 223-250

[Weihl 88] Weihl, W. E.; *Commutativity-based concurrency control for abstract data types*; IEEE Transactions On Computers, vol. 37, n° 12, December 1988, pp. 1488-1505

[Weikum 91] Weikum, G.; *Principles and realization strategies of multilevel transaction management*; ACM Transactions On Database Systems, vol. 16, n° 1, March 1991, pp. 132-180

[Yannakakis 84] Yannakakis, M.; *Serializability by locking*; Journal of the ACM, vol. 31, n° 2, April 1984, pp. 227-244